\documentclass[a4paper,11pt]{article}
\pdfoutput=1
\usepackage{jheppub}
\usepackage{amssymb,amsmath}

\usepackage[usenames,dvipsnames]{xcolor}
\usepackage{bm}
\usepackage{tikz}
\usetikzlibrary{intersections}
\usetikzlibrary{arrows,decorations.pathmorphing,backgrounds,positioning,fit,petri}
\usetikzlibrary{decorations.markings}
\usetikzlibrary{calc}
\usetikzlibrary{intersections,through,hobby}
\usetikzlibrary{external}
\definecolor{colA}{HTML}{c19277}
\definecolor{colB}{HTML}{e1bc91}
\definecolor{colD}{HTML}{62959c}

\usepackage{xargs} 
\usepackage[colorinlistoftodos,prependcaption,textsize=tiny]{todonotes}

\usepackage{blindtext}

\newcommand{\AH}[1]{\ensuremath{\mathcal{A}_{#1}}}
\newcommand{\SH}[1]{\ensuremath{\mathcal{S}_{#1}}}

\newcommand{\Li}{\mathrm{Li}}

\newcommand{\ep}{\ensuremath{\epsilon}}
\newcommand{\en}{\ensuremath{\mathcal{E}}}
\newcommand{\spe}{\ensuremath{\Sigma}}

\newcommand{\nmax}{16}

\title{Charge asymmetry in electron/positron energy loss in nuclear Bremsstrahlung.}

\author[a]{Roman N. Lee}
\author[b]{and Andrey F. Pikelner}

\affiliation[a]{Budker Institute of Nuclear Physics, Novosibirsk 630090, Russia}
\affiliation[b]{Joint Institute for Nuclear Research, Joliot-Curie, 6, Dubna 141980, Russia}

\emailAdd{r.n.lee@inp.nsk.su}
\emailAdd{pikelner@theor.jinr.ru}

\abstract{We calculate the leading Coulomb correction to the energy loss in the electron-nucleus Bremsstrahlung for arbitrary energy of the incoming particle. This correction determines the charge asymmetry, i.e., the difference of electron and positron energy loss. The result is presented in terms of the classical polylogarithms $\Li_2$ and $\Li_3$. We use modern multiloop methods based on the IBP reduction and on the differential equations for master integrals. We provide both the threshold and the high-energy asymptotics of the found asymmetry and compare them with the available results.}

\begin{document}

\maketitle
\flushbottom

\section{Introduction}
The electron Bremsstrahlung in nuclear field is one of the most fundamental processes in QED. 
Its investigation, both theoretical and experimental,  has a long history going back to Refs. 
\cite{heitler1933stopping,BetheHeitler1934,Racah1934}. 
In these papers, the spectrum and the radiation energy weighted cross section have been calculated in the leading Born approximation. 
The first two papers used the high-energy approximation $\en\gg m$, where $\en$ is the energy of the incoming electron and $m$ is the electron mass. 
Later on a lot of efforts has been devoted to the exact account of the Coulomb corrections and the screening at high energies, \cite{BethMax1954, OlseMax1959}. 
The first correction in $m/\en$ to the photon spectrum exact in the nucleus charge has been calculated in Ref. \cite{Lee2005}. 
This correction extended the applicability of the resulting formula for the spectrum  to the region of intermediate energies. 
However, it seems that, up to a present time, little was known about the Coulomb corrections in the region $\en-m\sim m$. 
In particular, the leading contribution to the charge asymmetry of the energy loss was never considered before in this energy range, to the best of our knowledge. 

The present paper fills this gap by calculating the first Coulomb correction to the energy-weighted cross section. 
To this end, we apply contemporary multiloop methods based on IBP reduction, differential equations and dimensional recurrence relations. Last but not least, we use this calculation to demonstrate a new technique of obtaining the dimensional recurrence relations for the boundary constants starting from dimensional recurrence relations and differential equations for the original master integrals which depend on parameter.

\section{Energy loss}

Energy loss in electron Bremsstrahlung on nucleus is defined as 
\begin{equation}\label{eq:phi}
    \phi = \int \frac{\omega}{\en} d\sigma_{eZ\to e\gamma Z} =\int \frac{\omega}{\en}\frac{\overline{|M|^2}}{2|\boldsymbol p|} \frac{d\Phi}{2\en}\,,\qquad d\Phi = 2\pi\delta(\en-\en'-\omega) \frac{d\boldsymbol{p}'}{2\en'(2\pi)^3} \frac{d\boldsymbol{k}}{2\omega(2\pi)^3}\,,
\end{equation}
where $(\en,\boldsymbol{p}),\ (\en',\boldsymbol{p}'),\ \text{and}\ (\omega,\boldsymbol{k})$ are the momenta of the incoming electron, scattered electron and emitted photon, respectively, and $\overline{|M|^2}$ denotes the square of matrix element averaged/summed over the polarizations of initial/final particles. The physical meaning of $\phi$ is the average fraction of electron energy lost per collision. This quantity plays an important role in the description of the particle propagation in the matter: when multiplied by the density of the scattering centers, it gives the \textit{stopping power} $S(\en)=-d\en/dx$, which in turn defines the \textit{radiation length} $X_0=\int_0^{\en} d\en_1/S(\en_1)$. It depends on the initial energy $\en$ and the nucleus charge number $Z$. 
The energy loss of the positron is given by a formal replacement $Z\to -Z$.
When $Z\alpha\ll 1$ (here $Z$ is the charge number of the nucleus), $\phi$ can be expanded in  $Z\alpha$ with the leading term $\propto \alpha(Z\alpha)^2$. 
Energy loss in the leading Born approximation has been calculated long ago in Racah paper \cite{Racah1934}: 
\begin{align}
    \label{eq:rphi-born-mEdep}
  \bar\phi_{\mathrm{LO}} = \frac{12\en^2+4m^2}{3 \en p } \ln{\frac{\en+p}{m}}
                       - \frac{8\en+6p }{3 \en p^2}\ln^2\frac{\en+p}{m} -\frac{4}{3}
                       - \frac{2}{\en p}\Li_2\left( -\frac{2p(\en+p)}{m^2} \right)\,,
\end{align}
where $\bar\phi \stackrel{\text{def}}{=} \phi/\sigma_0$ with $\sigma_0 \stackrel{\text{def}}{=}\alpha(Z\alpha)^2/m^2$.

In order to calculate the charge asymmetry in the leading approximation, one has to account for the first correction with respect to the parameter $Z\alpha$.
Let $M=\sum_{k=1}^{\infty} M_k$, where the term $M_k\propto (Z\alpha)^k$ corresponds to the contribution of diagrams with $k$ Coulomb exchanges between electron and nucleus. 
Then the first $C$-odd correction to $\phi$ is determined by Eq. 
\eqref{eq:phi} with the replacement $\left|M\right|^2\to 2\mathrm{Re}[M_{2}^{*}M_{1}]$.

\section{Methods} 
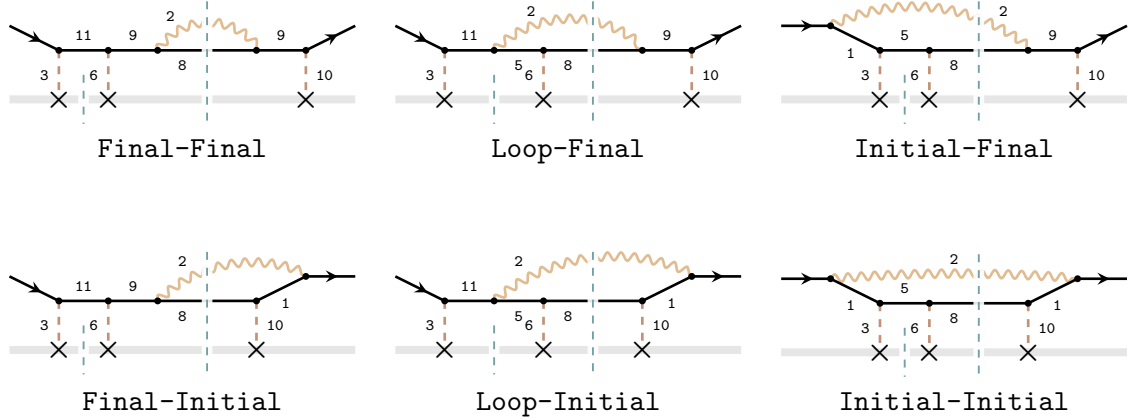
\begin{figure}
  \centering
    \tikzset{
    	di/.style={line width=1pt,draw=black, postaction={decorate},
    		decoration={markings,mark=at position .65 with
    			{\arrow[scale=1,draw=black,>=stealth]{>}}}},
    	ndi/.style={line width=1pt,draw=black},
    	photon/.style={draw=colB,line width=1pt,decorate,
    		decoration={snake,aspect=0.5,amplitude=1.5pt,segment
    			length=5pt}},
    	ext/.style={line width=1pt,draw=colA,dashed},
    	nu/.style={line width=3pt,draw=gray!20!white},
    	cutbg/.style={line width=4pt,opacity=0.9,draw=white},
    	cut/.style={line width=0.7pt,dashed, draw=colD!90!white}
    }
    
    \begin{tabular}{ccc}
    	\begin{tikzpicture}[use Hobby shortcut,baseline={(current bounding box.center)},scale=0.65]
    		\coordinate (vIN) at (-4,0.5);
    		\coordinate (vOUT) at (3,0.5);
    		\coordinate (vf0) at (0,0);
    		%
    		\coordinate (vLIf1) at (-2,0);
    		\coordinate (vLIf2) at (-1,0);
    		\coordinate (vLIn1) at (-2,-1);
    		\coordinate (vLIn2) at (-1,-1);
    		\coordinate (vLIg) at (-3,0.5);
    		\coordinate (vLLf1) at (-3,0);
    		\coordinate (vLLf2) at (-1,0);
    		\coordinate (vLLn1) at (-3,-1);
    		\coordinate (vLLn2) at (-1,-1);
    		\coordinate (vLLg) at (-2,0);
    		\coordinate (vLFf1) at (-3,0);
    		\coordinate (vLFf2) at (-2,0);
    		\coordinate (vLFn1) at (-3,-1);
    		\coordinate (vLFn2) at (-2,-1);
    		\coordinate (vLFg) at (-1,0);
    		%
    		\coordinate (vRIf) at (1,0);
    		\coordinate (vRIn) at (1,-1);
    		\coordinate (vRIg) at (2,0.5);
    		\coordinate (vRFf) at (2,0);
    		\coordinate (vRFn) at (2,-1);
    		\coordinate (vRFg) at (1,0);
    		%
    		%
    		%
    		\draw[nu] (-4,-1) -- (3,-1);
    		\draw[photon] (vLFg) .. 
    		(-0.2,0.5) .. 
    		(vRFg);
    		%
    		%
    		%
    		\draw[di] (vIN) -- (vLFf1);
    		\draw[ndi] (vLFf1) -- (vLFf2);
    		\draw[ndi] (vLFf2) -- (vLFg);
    		\draw[ndi] (vLFg) -- (vf0);
    		\draw[ext] (vLFf1) -- (vLFn1);
    		\draw[ext] (vLFf2) -- (vLFn2);
    		\draw (vLFn1) node[anchor=center] {$\bm{\times}$};
    		\draw (vLFn2) node[anchor=center] {$\bm{\times}$};
    		\fill (vLFg) circle (2pt);
    		\fill (vLFf1) circle (2pt);
    		\fill (vLFf2) circle (2pt);
    		\draw[ndi] (vf0) -- (vRFg);
    		\draw[ndi] (vRFg) -- (vRFf);
    		\draw[di] (vRFf) -- (vOUT);
    		\draw[ext] (vRFf) -- (vRFn);
    		\draw (vRFn) node[anchor=center] {$\bm{\times}$};
    		\fill (vRFf) circle (2pt);
    		\fill (vRFg) circle (2pt);
    		\draw[cutbg] (0,1) -- (0,-1.5);
    		\draw[cut] (0,1) -- (0,-1.5);
    		\draw[cutbg] (-2.5,-0.5) -- (-2.5,-1.5);
    		\draw[cut] (-2.5,-0.5) -- (-2.5,-1.5);
    		\draw (-3,-0.5) node[anchor=east] {\tiny \texttt{3}};
    		\draw (-2,-0.5) node[anchor=east] {\tiny \texttt{6}};
    		\draw (2,-0.5) node[anchor=west] {\tiny \texttt{10}};
    		\draw (-2.5,0) node[anchor=south] {\tiny \texttt{11}};
    		\draw (-1.5,0) node[anchor=south] {\tiny \texttt{9}};
    		\draw (-0.5,0) node[anchor=north] {\tiny \texttt{8}};
    		\draw (1.5,0) node[anchor=south] {\tiny \texttt{9}};
    		\draw (-0.74,0.4) node[anchor=south] {\tiny \texttt{2}};
    	\end{tikzpicture}
    	&
    	\begin{tikzpicture}[use Hobby shortcut,baseline={(current bounding box.center)},scale=0.65]
    		\coordinate (vIN) at (-4,0.5);
    		\coordinate (vOUT) at (3,0.5);
    		\coordinate (vf0) at (0,0);
    		%
    		\coordinate (vLIf1) at (-2,0);
    		\coordinate (vLIf2) at (-1,0);
    		\coordinate (vLIn1) at (-2,-1);
    		\coordinate (vLIn2) at (-1,-1);
    		\coordinate (vLIg) at (-3,0.5);
    		\coordinate (vLLf1) at (-3,0);
    		\coordinate (vLLf2) at (-1,0);
    		\coordinate (vLLn1) at (-3,-1);
    		\coordinate (vLLn2) at (-1,-1);
    		\coordinate (vLLg) at (-2,0);
    		\coordinate (vLFf1) at (-3,0);
    		\coordinate (vLFf2) at (-2,0);
    		\coordinate (vLFn1) at (-3,-1);
    		\coordinate (vLFn2) at (-2,-1);
    		\coordinate (vLFg) at (-1,0);
    		%
    		\coordinate (vRIf) at (1,0);
    		\coordinate (vRIn) at (1,-1);
    		\coordinate (vRIg) at (2,0.5);
    		\coordinate (vRFf) at (2,0);
    		\coordinate (vRFn) at (2,-1);
    		\coordinate (vRFg) at (1,0);
    		%
    		%
    		%
    		\draw[nu] (-4,-1) -- (3,-1);
    		\draw[photon] (vLLg) .. 
    		(-0.5,0.6) .. 
    		(vRFg);
    		%
    		%
    		%
    		\draw[di] (vIN) -- (vLLf1);
    		\draw[ndi] (vLLf1) -- (vLLg);
    		\draw[ndi] (vLLg) -- (vLLf2);
    		\draw[ndi] (vLLf2) -- (vf0);
    		\draw[ext] (vLLf1) -- (vLLn1);
    		\draw[ext] (vLLf2) -- (vLLn2);
    		\draw (vLLn1) node[anchor=center] {$\bm{\times}$};
    		\draw (vLLn2) node[anchor=center] {$\bm{\times}$};
    		\fill (vLLg) circle (2pt);
    		\fill (vLLf1) circle (2pt);
    		\fill (vLLf2) circle (2pt);
    		\draw[ndi] (vf0) -- (vRFg);
    		\draw[ndi] (vRFg) -- (vRFf);
    		\draw[di] (vRFf) -- (vOUT);
    		\draw[ext] (vRFf) -- (vRFn);
    		\draw (vRFn) node[anchor=center] {$\bm{\times}$};
    		\fill (vRFf) circle (2pt);
    		\fill (vRFg) circle (2pt);
    		\draw[cutbg] (0,1) -- (0,-1.5);
    		\draw[cut] (0,1) -- (0,-1.5);
    		\draw[cutbg] (-2,-0.5) -- (-2,-1.5);
    		\draw[cut] (-2,-0.5) -- (-2,-1.5);
    		\draw (-3,-0.5) node[anchor=east] {\tiny \texttt{3}};
    		\draw (-1,-0.5) node[anchor=east] {\tiny \texttt{6}};
    		\draw (2,-0.5) node[anchor=west] {\tiny \texttt{10}};
    		\draw (-2.5,0) node[anchor=south] {\tiny \texttt{11}};
    		\draw (-1.5,0) node[anchor=north] {\tiny \texttt{5}};
    		\draw (1.5,0) node[anchor=south] {\tiny \texttt{9}};
    		\draw (-0.5,0) node[anchor=north] {\tiny \texttt{8}};
    		\draw (-1.25,0.5) node[anchor=south] {\tiny \texttt{2}};
    	\end{tikzpicture}
    	&
    	\begin{tikzpicture}[use Hobby shortcut,baseline={(current bounding box.center)},scale=0.65]
    		\coordinate (vIN) at (-4,0.5);
    		\coordinate (vOUT) at (3,0.5);
    		\coordinate (vf0) at (0,0);
    		%
    		\coordinate (vLIf1) at (-2,0);
    		\coordinate (vLIf2) at (-1,0);
    		\coordinate (vLIn1) at (-2,-1);
    		\coordinate (vLIn2) at (-1,-1);
    		\coordinate (vLIg) at (-3,0.5);
    		\coordinate (vLLf1) at (-3,0);
    		\coordinate (vLLf2) at (-1,0);
    		\coordinate (vLLn1) at (-3,-1);
    		\coordinate (vLLn2) at (-1,-1);
    		\coordinate (vLLg) at (-2,0);
    		\coordinate (vLFf1) at (-3,0);
    		\coordinate (vLFf2) at (-2,0);
    		\coordinate (vLFn1) at (-3,-1);
    		\coordinate (vLFn2) at (-2,-1);
    		\coordinate (vLFg) at (-1,0);
    		%
    		\coordinate (vRIf) at (1,0);
    		\coordinate (vRIn) at (1,-1);
    		\coordinate (vRIg) at (2,0.5);
    		\coordinate (vRFf) at (2,0);
    		\coordinate (vRFn) at (2,-1);
    		\coordinate (vRFg) at (1,0);
    		%
    		%
    		%
    		\draw[nu] (-4,-1) -- (3,-1);
    		\draw[photon] (vLIg) .. 
    		(0,0.6) .. 
    		(vRFg);
    		%
    		%
    		%
    		\draw[di] (vIN) -- (vLIg);
    		\draw[ndi] (vLIg) -- (vLIf1);
    		\draw[ndi] (vLIf1) -- (vLIf2);
    		\draw[ndi] (vLIf2) -- (vf0);
    		\draw[ext] (vLIf1) -- (vLIn1);
    		\draw[ext] (vLIf2) -- (vLIn2);
    		\draw (vLIn1) node[anchor=center] {$\bm{\times}$};
    		\draw (vLIn2) node[anchor=center] {$\bm{\times}$};
    		\fill (vLIg) circle (2pt);
    		\fill (vLIf1) circle (2pt);
    		\fill (vLIf2) circle (2pt);
    		\draw[ndi] (vf0) -- (vRFg);
    		\draw[ndi] (vRFg) -- (vRFf);
    		\draw[di] (vRFf) -- (vOUT);
    		\draw[ext] (vRFf) -- (vRFn);
    		\draw (vRFn) node[anchor=center] {$\bm{\times}$};
    		\fill (vRFf) circle (2pt);
    		\fill (vRFg) circle (2pt);
    		\draw[cutbg] (0,1) -- (0,-1.5);
    		\draw[cut] (0,1) -- (0,-1.5);
    		\draw[cutbg] (-1.5,-0.5) -- (-1.5,-1.5);
    		\draw[cut] (-1.5,-0.5) -- (-1.5,-1.5);
    		\draw (-2,-0.5) node[anchor=east] {\tiny \texttt{3}};
    		\draw (-1,-0.5) node[anchor=east] {\tiny \texttt{6}};
    		\draw (2,-0.5) node[anchor=west] {\tiny \texttt{10}};
    		\draw (-2.3,0.25) node[anchor=north east] {\tiny \texttt{1}};
    		\draw (-1.5,0) node[anchor=south] {\tiny \texttt{5}};
    		\draw (1.5,0) node[anchor=south] {\tiny \texttt{9}};
    		\draw (-0.5,0) node[anchor=north] {\tiny \texttt{8}};
    		\draw (0.5,0.5) node[anchor=south] {\tiny \texttt{2}};
    	\end{tikzpicture}
    	\\
    	\texttt{Final-Final} & \texttt{Loop-Final} & \texttt{Initial-Final}\vspace{1cm}\\
    	\begin{tikzpicture}[use Hobby shortcut,baseline={(current bounding box.center)},scale=0.65]
    		\coordinate (vIN) at (-4,0.5);
    		\coordinate (vOUT) at (3,0.5);
    		\coordinate (vf0) at (0,0);
    		%
    		\coordinate (vLIf1) at (-2,0);
    		\coordinate (vLIf2) at (-1,0);
    		\coordinate (vLIn1) at (-2,-1);
    		\coordinate (vLIn2) at (-1,-1);
    		\coordinate (vLIg) at (-3,0.5);
    		\coordinate (vLLf1) at (-3,0);
    		\coordinate (vLLf2) at (-1,0);
    		\coordinate (vLLn1) at (-3,-1);
    		\coordinate (vLLn2) at (-1,-1);
    		\coordinate (vLLg) at (-2,0);
    		\coordinate (vLFf1) at (-3,0);
    		\coordinate (vLFf2) at (-2,0);
    		\coordinate (vLFn1) at (-3,-1);
    		\coordinate (vLFn2) at (-2,-1);
    		\coordinate (vLFg) at (-1,0);
    		%
    		\coordinate (vRIf) at (1,0);
    		\coordinate (vRIn) at (1,-1);
    		\coordinate (vRIg) at (2,0.5);
    		\coordinate (vRFf) at (2,0);
    		\coordinate (vRFn) at (2,-1);
    		\coordinate (vRFg) at (1,0);
    		%
    		%
    		%
    		\draw[nu] (-4,-1) -- (3,-1);
    		\draw[photon] (vLFg) .. 
    		(0,0.6) .. 
    		(vRIg);
    		%
    		%
    		%
    		\draw[di] (vIN) -- (vLFf1);
    		\draw[ndi] (vLFf1) -- (vLFf2);
    		\draw[ndi] (vLFf2) -- (vLFg);
    		\draw[ndi] (vLFg) -- (vf0);
    		\draw[ext] (vLFf1) -- (vLFn1);
    		\draw[ext] (vLFf2) -- (vLFn2);
    		\draw (vLFn1) node[anchor=center] {$\bm{\times}$};
    		\draw (vLFn2) node[anchor=center] {$\bm{\times}$};
    		\fill (vLFg) circle (2pt);
    		\fill (vLFf1) circle (2pt);
    		\fill (vLFf2) circle (2pt);
    		\draw[ndi] (vf0) -- (vRIf);
    		\draw[ndi] (vRIf) -- (vRIg);
    		\draw[di] (vRIg) -- (vOUT);
    		\draw[ext] (vRIf) -- (vRIn);
    		\draw (vRIn) node[anchor=center] {$\bm{\times}$};
    		\fill (vRIf) circle (2pt);
    		\fill (vRIg) circle (2pt);
    		\draw[cutbg] (0,1) -- (0,-1.5);
    		\draw[cut] (0,1) -- (0,-1.5);
    		\draw[cutbg] (-2.5,-0.5) -- (-2.5,-1.5);
    		\draw[cut] (-2.5,-0.5) -- (-2.5,-1.5);
    		\draw (-3,-0.5) node[anchor=east] {\tiny \texttt{3}};
    		\draw (-2,-0.5) node[anchor=east] {\tiny \texttt{6}};
    		\draw (1,-0.5) node[anchor=west] {\tiny \texttt{10}};
    		\draw (-2.5,0) node[anchor=south] {\tiny \texttt{11}};
    		\draw (-1.5,0) node[anchor=south] {\tiny \texttt{9}};
    		\draw (1.3,0.26) node[anchor=north west] {\tiny \texttt{1}};
    		\draw (-0.5,0) node[anchor=north] {\tiny \texttt{8}};
    		\draw (-0.5,0.5) node[anchor=south] {\tiny \texttt{2}};
    	\end{tikzpicture}
    	&
    	\begin{tikzpicture}[use Hobby shortcut,baseline={(current bounding box.center)},scale=0.65]
    		\coordinate (vIN) at (-4,0.5);
    		\coordinate (vOUT) at (3,0.5);
    		\coordinate (vf0) at (0,0);
    		%
    		\coordinate (vLIf1) at (-2,0);
    		\coordinate (vLIf2) at (-1,0);
    		\coordinate (vLIn1) at (-2,-1);
    		\coordinate (vLIn2) at (-1,-1);
    		\coordinate (vLIg) at (-3,0.5);
    		\coordinate (vLLf1) at (-3,0);
    		\coordinate (vLLf2) at (-1,0);
    		\coordinate (vLLn1) at (-3,-1);
    		\coordinate (vLLn2) at (-1,-1);
    		\coordinate (vLLg) at (-2,0);
    		\coordinate (vLFf1) at (-3,0);
    		\coordinate (vLFf2) at (-2,0);
    		\coordinate (vLFn1) at (-3,-1);
    		\coordinate (vLFn2) at (-2,-1);
    		\coordinate (vLFg) at (-1,0);
    		%
    		\coordinate (vRIf) at (1,0);
    		\coordinate (vRIn) at (1,-1);
    		\coordinate (vRIg) at (2,0.5);
    		\coordinate (vRFf) at (2,0);
    		\coordinate (vRFn) at (2,-1);
    		\coordinate (vRFg) at (1,0);
    		%
    		%
    		%
    		\draw[nu] (-4,-1) -- (3,-1);
    		\draw[photon] (vLLg) .. 
    		(-1,0.6) .. 
    		(vRIg);
    		%
    		%
    		%
    		\draw[di] (vIN) -- (vLLf1);
    		\draw[ndi] (vLLf1) -- (vLLg);
    		\draw[ndi] (vLLg) -- (vLLf2);
    		\draw[ndi] (vLLf2) -- (vf0);
    		\draw[ext] (vLLf1) -- (vLLn1);
    		\draw[ext] (vLLf2) -- (vLLn2);
    		\draw (vLLn1) node[anchor=center] {$\bm{\times}$};
    		\draw (vLLn2) node[anchor=center] {$\bm{\times}$};
    		\fill (vLLg) circle (2pt);
    		\fill (vLLf1) circle (2pt);
    		\fill (vLLf2) circle (2pt);
    		\draw[ndi] (vf0) -- (vRIf);
    		\draw[ndi] (vRIf) -- (vRIg);
    		\draw[di] (vRIg) -- (vOUT);
    		\draw[ext] (vRIf) -- (vRIn);
    		\draw (vRIn) node[anchor=center] {$\bm{\times}$};
    		\fill (vRIf) circle (2pt);
    		\fill (vRIg) circle (2pt);
    		\draw[cutbg] (0,1) -- (0,-1.5);
    		\draw[cut] (0,1) -- (0,-1.5);
    		\draw[cutbg] (-2,-0.5) -- (-2,-1.5);
    		\draw[cut] (-2,-0.5) -- (-2,-1.5);
    		\draw (-3,-0.5) node[anchor=east] {\tiny \texttt{3}};
    		\draw (-1,-0.5) node[anchor=east] {\tiny \texttt{6}};
    		\draw (1,-0.5) node[anchor=west] {\tiny \texttt{10}};
    		\draw (-2.5,0) node[anchor=south] {\tiny \texttt{11}};
    		\draw (-1.5,0) node[anchor=north] {\tiny \texttt{5}};
    		\draw (1.3,0.26) node[anchor=north west] {\tiny \texttt{1}};
    		\draw (-0.5,0) node[anchor=north] {\tiny \texttt{8}};
    		\draw (-1.5,0.5) node[anchor=south] {\tiny \texttt{2}};
    	\end{tikzpicture}
    	&
    	\begin{tikzpicture}[use Hobby shortcut,baseline={(current bounding box.center)},scale=0.65]
    		\coordinate (vIN) at (-4,0.5);
    		\coordinate (vOUT) at (3,0.5);
    		\coordinate (vf0) at (0,0);
    		%
    		\coordinate (vLIf1) at (-2,0);
    		\coordinate (vLIf2) at (-1,0);
    		\coordinate (vLIn1) at (-2,-1);
    		\coordinate (vLIn2) at (-1,-1);
    		\coordinate (vLIg) at (-3,0.5);
    		\coordinate (vLLf1) at (-3,0);
    		\coordinate (vLLf2) at (-1,0);
    		\coordinate (vLLn1) at (-3,-1);
    		\coordinate (vLLn2) at (-1,-1);
    		\coordinate (vLLg) at (-2,0);
    		\coordinate (vLFf1) at (-3,0);
    		\coordinate (vLFf2) at (-2,0);
    		\coordinate (vLFn1) at (-3,-1);
    		\coordinate (vLFn2) at (-2,-1);
    		\coordinate (vLFg) at (-1,0);
    		%
    		\coordinate (vRIf) at (1,0);
    		\coordinate (vRIn) at (1,-1);
    		\coordinate (vRIg) at (2,0.5);
    		\coordinate (vRFf) at (2,0);
    		\coordinate (vRFn) at (2,-1);
    		\coordinate (vRFg) at (1,0);
    		%
    		%
    		%
    		\draw[nu] (-4,-1) -- (3,-1);
    		\draw[photon] (vLIg) .. 
    		(-0.5,0.6) .. 
    		(vRIg);
    		%
    		%
    		%
    		\draw[di] (vIN) -- (vLIg);
    		\draw[ndi] (vLIg) -- (vLIf1);
    		\draw[ndi] (vLIf1) -- (vLIf2);
    		\draw[ndi] (vLIf2) -- (vf0);
    		\draw[ext] (vLIf1) -- (vLIn1);
    		\draw[ext] (vLIf2) -- (vLIn2);
    		\draw (vLIn1) node[anchor=center] {$\bm{\times}$};
    		\draw (vLIn2) node[anchor=center] {$\bm{\times}$};
    		\fill (vLIg) circle (2pt);
    		\fill (vLIf1) circle (2pt);
    		\fill (vLIf2) circle (2pt);
    		\draw[ndi] (vf0) -- (vRIf);
    		\draw[ndi] (vRIf) -- (vRIg);
    		\draw[di] (vRIg) -- (vOUT);
    		\draw[ext] (vRIf) -- (vRIn);
    		\draw (vRIn) node[anchor=center] {$\bm{\times}$};
    		\fill (vRIf) circle (2pt);
    		\fill (vRIg) circle (2pt);
    		\draw[cutbg] (0,1) -- (0,-1.5);
    		\draw[cut] (0,1) -- (0,-1.5);
    		\draw[cutbg] (-1.5,-0.5) -- (-1.5,-1.5);
    		\draw[cut] (-1.5,-0.5) -- (-1.5,-1.5);
    		\draw (-2,-0.5) node[anchor=east] {\tiny \texttt{3}};
    		\draw (-1,-0.5) node[anchor=east] {\tiny \texttt{6}};
    		\draw (1,-0.5) node[anchor=west] {\tiny \texttt{10}};
    		\draw (-2.3,0.26) node[anchor=north east] {\tiny \texttt{1}};
    		\draw (-1.5,0) node[anchor=south] {\tiny \texttt{5}};
    		\draw (1.3,0.26) node[anchor=north west] {\tiny \texttt{1}};
    		\draw (-0.5,0) node[anchor=north] {\tiny \texttt{8}};
    		\draw (-0.5,0.6) node[anchor=south] {\tiny \texttt{2}};
    	\end{tikzpicture}
    	\\
    	\texttt{Final-Initial} & \texttt{Loop-Initial} & \texttt{Initial-Initial}\vspace{5mm}\\
    \end{tabular}
    \caption{Diagrams contributing to NLO cross section}
    \label{fig:diagrams}
\end{figure}
In order to apply multiloop methods, we use dimensional regularization and express the phase-spaces integrals via loop integrals with cut propagator using relation
\begin{equation}
    \frac{d^{d-1}\boldsymbol{p}'}{2\en'(2\pi)^{d-1}}=\frac{d^d p'}{(2\pi)^d}2\pi\delta_+(p'^2-m^2)
\end{equation}

Then the contribution of $M_{2}^{*}M_{1}$ is given by the sum of cut diagrams
depicted in Fig. \ref{fig:diagrams}. In the limit $\ep\to 0$ the quantity $M_2$
is infrared divergent due to long range of the Coulomb interaction. However,
this divergence is known to be absorbed into a complex phase factor in the full amplitude. It means that $M_{2}=\frac
{1}\ep aM_1+O(\ep^0)$ where $a$ is a purely imaginary number, so that
$\mathrm{Re}[M_{2}^{*}M_{1}]$ has finite limit $\ep\to0$.
Then, the integration
over final particles phase space also has a finite limit $\ep\to 0$  as the infrared divergencies do not show up due to additional factor $\omega$ under the integral sign.

We consider the following family of integrals

\begin{equation}
  \label{eq:jIntPows}
  j\left(n_{1},\ldots,n_{12}\right)=\int\frac{dqdkdp'}{i\pi^{3d/2}}
    \prod_{k\in \{1,\ldots,12\}\backslash C}\left(D_{k}+i0\right)^{-n_{k}}\times
    \prod_{k\in C}\frac{\delta^{(n_k-1)}\left(-D_{k}\right)}{(n_k-1)!}\,,
\end{equation}
where the set $C=\left\{ 2,4,7,8\right\} $ enumerates the cut denominators
\begin{gather*}
  D_{2}=k^{2}, \qquad D_{4}=q\cdot n, \qquad D_{7}=Q\cdot n,\qquad  D_{8}=p'^2.
  \end{gather*}
  where $Q=p-p'-k$ and $n=(1,\boldsymbol 0)$ is the time ort. These cut denominators correspond to the on-shell condition for the emitted photon $(2)$, the zero energy transfer to the heavy nucleus $(4,7)$ and the on-shell condition for the final electron $(8)$. The remaining propagators are
  \begin{align*}
    &D_{1}  =\left(p-k\right){}^{2}-m^2\,,  &  &D_{3}  =q^{2}\,,&  &D_{5}  =\left(p-k-q\right){}^{2}-m^2\,,&  &D_{6} =\left(Q-q\right){}^{2}\,, \\
    &D_{9}  =\left(k+p'\right){}^{2}-m^2\,, &  &D_{10} =Q^{2}\,,&  &D_{11} =\left(p-q\right){}^{2}-m^2\,,  &  &D_{12} =k\cdot n\,.
  \end{align*}


All diagrams in Fig. \ref{fig:diagrams} are expressed via integrals $j(n_1,\ldots n_{12})$ for which $n_{k\in C}>0$, $n_{12}\leqslant 0$ and at least one of $n_1,\ n_5,\ n_9,\ n_{11}$ is non-positive.

IBP reduction\footnote{IBP reduction was performed using the \texttt{LiteRed2} package \cite{LiteRed2012,LiteRed2013,LiteRed2021}.} reveals 61 master integrals. We construct differential system 
\begin{equation}
    \partial_\en \bm j = M \bm j,
\end{equation}
where $\bm j=(j_1,\ldots j_{61})^\intercal$ is a column of \texttt{LiteRed} master integrals. Then we reduce this system to $\ep$-form \cite{Henn2013,Lee2014} using \texttt{Libra} package \cite{Libra2020}. In order to find the transformation 
\begin{equation}\label{eq:jtoJ}
    \bm j=T\bm J
\end{equation} to the canonical basis $\bm J = (J_1,\ldots, J_{61})$, we pass to
the variable $z\in(0,1)$ related to $\en$ via
\begin{equation}
    \en=m\frac{1+z^2}{1-z^2}\,,\qquad z=\sqrt{\frac{\en-m}{\en+m}}.
\end{equation} 
Then we have
\begin{equation}\label{eq:e-form}
    \partial_z \bm J=\ep A(z)\bm J=\ep\left[ \frac{A_0}{z}+\frac{A_1}{z-1}+\frac{A_2}{z+1}+\frac{2z A_3}{z^2+1}\right]\bm J\,,
\end{equation}
where $A_n$ are some constant matrices.

\subsection*{Boundary conditions}
We represent the solution as
\begin{equation}\label{eq:ULCs}
    \bm J (z) 
    = U(z) \bm C \,,
\end{equation} 
where $U(z)$ 
is the evolution operator and $ \bm C$ is a column of the boundary constants to be fixed. We fix the asymptotics of the evolution operator to be $U(z)\sim z^{\ep A_0}$, where $A_0$ is defined in Eq. \eqref{eq:e-form}. Then, around the point  $z=0$, the evolution operator $U(z)$ can be expanded in the generalized power series of the form
\begin{multline}\label{eq:Uexpansion}
    U(z) = \sum_{n=0}^{\infty}\bigg\{U_{n} z^n+U_{n+2\ep} z^{n+2\ep}
    +U_{n-2\ep} z^{n-2\ep}+U_{n-4\ep} z^{n-4\ep}+U^{(1)}_{n-4\ep} z^{n-4\ep}\log z\\
    +U_{n-6\ep} z^{n-6\ep}+U_{n-8\ep} z^{n-8\ep}\bigg\}\,,
\end{multline}
where the leading terms ($n=0$) are simply given by $z^{\ep A_0}$ and next terms can be found from recurrence relation,  as described in Ref. \cite{Lee:2017qql}. From this expansion and Eq. \eqref{eq:jtoJ}, it is obvious that the  boundary constants $ \bm C $ can be determined by fixing a properly chosen set of 61 asymptotic coefficients of master integrals. In order to determine which specific set is suitable, we use the approach described in Ref. \cite{Lee:2019zop}. \texttt{Libra} has all necessary tools for finding a column of asymptotic coefficients $\bm c$ required to fix the boundary conditions and the `adapter' matrix $L$ which maps this column to the column of boundary constants, such that $\bm C= L\bm c$ and
\begin{equation}\label{eq:ULcs}
    \bm J (z)  = U(z) L \bm c\,.
\end{equation} 
For the description of these tools, we would refer the reader to Ref. \cite{Libra2020}. Using \texttt{Libra}, we discover that it is sufficient to determine
\begin{itemize}
    \item Coefficients in front of $z^{6-8\ep},z^{6-8\ep},z^{4-8\ep},z^{4-8\ep},z^{4-8\ep},z^{2-8\ep}$ for the master integrals \#\# 1, 3, 9, 16, 19, 28, respectively;
    \item 12 coefficients in front of $z^{n-6\ep}$ ($n\in \mathbb{Z}$);
    \item 43 coefficients in total in front of $z^{n-4\ep}$, $z^{n-2\ep}$, $z^{n}$ or $z^{n+2\ep}$  ($n\in \mathbb{Z}$).
\end{itemize}


Let us explain how we obtain the dimensional recurrence relations
for the column of boundary constants. We construct dimensional recurrence
relation for the \texttt{LiteRed} master
integrals
\begin{equation}
    \bm j(z,\ep-1)=R(z,\ep)\bm j(z,\ep)\,, 
\end{equation}
where $R(z,\ep)$ is some rational matrix. Now from Eqs. \eqref{eq:ULcs} and \eqref{eq:jtoJ} we get
\begin{equation}
    \bm j=\widetilde{T}\bm c\,,\qquad \widetilde{T}=T\,U(z) L, 
\end{equation}
and treat $\widetilde{T}$ as a transformation matrix connecting $\bm j$ and the column of coefficients $\bm c$. Then the dimensional recurrence relation for $\bm c$ is given by 
\begin{equation}\label{eq:cDRR}
    \bm c (\epsilon -1)=V(\ep)\bm c(\epsilon)\,,
\end{equation}
where
\begin{equation}\label{eq:V}
    V(\ep)=\widetilde{T}^{-1}(z,\ep-1)R(z,\epsilon )\widetilde{T}\left(z,\epsilon \right)\,.
\end{equation}
Note that the left-hand side of Eq. \eqref{eq:V} does not depend on $z$, so the
right-hand side should not depend on $z$ also. It is essential that
\texttt{Libra} has tools for expanding both
$\widetilde{T}^{-1}\left(\epsilon -1\right)$ and $\widetilde{T}\left(\epsilon \right)$ in generalized power series in $z$. Thus, expanding the right-hand side of Eq. \eqref{eq:V} up to sufficiently high order, we find the exact matrix $V$ entering the dimensional recurrence relation \eqref{eq:cDRR}. Note that the required depth of the expansion of $U$ in $z$ appears to be as high as $11$ since the powers
$z^{n-8\ep}$ related to the nonzero coefficients in $\bm c$ column change
essentially upon the shift $\ep\to\ep-1$. Note that the matrix $V$ acquires a
block-diagonal form with each block corresponding to a specific $k$ in
$z^{n-k\ep}$ constants.
To summarize the above consideration, we have constructed the exact dimensional recurrence \eqref{eq:cDRR} for the asymptotic coefficients $\bm c$ starting from the dimensional recurrence for the master integrals $\bm j$ and the generalized power series expansion of the evolution operator $U(z)$, which, in turn, is constructed starting from the differential system \eqref{eq:e-form}. This approach is quite general and can be used in more involved calculations.

Now we note, that the phase space integration is subject to severe restrictions due to $\delta$-functions. As we shall see in a moment, this greatly reduces the number of  non-zero coefficients from the above list. We have the following power counting:
\[
p'\sim p\sim z,\qquad\omega\sim \en-m \sim z^{2}
\]
Thus, we have 
\[
d\boldsymbol{p}'d\boldsymbol{k}\sim z^{d-1}\left(z^{2}\right)^{d-1}=z^{9-6\epsilon},
\]
i.e. only powers $z^{n-k\epsilon}$ with $n\in\mathbb{Z},\ k\geqslant6$ survive in the threshold asymptotics. Moreover, the terms $\propto z^{n-6\epsilon}$ come from the region $|\boldsymbol{q}|\sim z^0$ and thorough inspection of the functions $D_k$ shows that the contribution of this region vanishes. Therefore, we only have to evaluate the six constants
\begin{equation}
    c_1^{6},\quad
    c_3^{6},\quad
    c_9^{4},\quad
    c_{16}^{4},\quad
    c_{19}^{4},\quad
    c_{28}^{2},\label{eq:cs}
\end{equation}
where $c_k^{n}$ denotes the coefficient in front of $z^{n-8\ep}$ in the
threshold asymptotics of $k$-th master integral. The first five constants can be
calculated exactly in $\ep$ using parametric representation in terms of either
product of $\Gamma$-functions or hypergeometric functions. However, as the
sixth constant, $c_{28}^{2}$, requires a special treatment anyway, we
prefer to use the DRA method \cite{DRA2010} for all constants in Eq.
\eqref{eq:cs}.

The DRA method requires, in addition to the dimensional recurrence relations
\eqref{eq:cDRR}, integral representations from which one can determine
analytical properties of $\bm c$ as functions of $\ep$. We use the standard
expansion by regions method to obtain the integral representations for each nonzero
constant. For the most complicated constant $c_{28}^2$ we obtain a three-fold
integral representation. Using \texttt{SummerTime} \cite{SummerTime2016} and PSLQ algorithm \cite{PSLQ}, we obtain the expansions of all six constants from Eq. \eqref{eq:cs} in terms of alternating multiple zeta values.

\section{Results}

%
%
%
We organize series expansion in $Z\alpha$ for the energy-loss in the following form
\begin{equation}
  \label{eq:phiLO-NLO-def}
  \phi = \sigma_0\bar{\phi} = 
  \sigma_0\left[ \bar{\phi}_{\rm LO} + (Z\alpha)
    \bar{\phi}_{\rm NLO} + \mathcal{O}\left(Z\alpha\right)^2\right]\,.
\end{equation}
Here $\bar{\phi}_{\rm LO}$ is given by Eq.~\eqref{eq:rphi-born-mEdep} and our new result for the NLO correction reads\footnote{One can find results for the $\bar{\phi}_{\rm NLO}$ in computer-readable form as the ancillary files to the \texttt{arXiv} submission.}
\begin{align}
  \label{eq:asyNLOAS}
  \bar{\phi}_{\rm NLO} & = \frac{\pi(1 - z^2)}{(1 + z^2)} \left(
                        \frac{(40 - 7 z^2 + 3 z^4)}{ 12 z} - \frac{84 - 19 z^2 + 71 z^4 + 59 z^6 - 3 z^8}{ 24 z^2 (1 - z^2)} \AH{1}\right.\nonumber\\
                       & + \frac{(3 + z^2) (1 + 3 z^2) (1 + 10 z^2 + z^4)}{24 z^3 (1 - z^2)} \left(2 \log{2} + \SH{1}\right)
                        + \frac{(1 - z^4) (1 + 10 z^2 + z^4)}{16 z^4} \AH{1, -1} 
                        \nonumber\\
                       &
                       + \frac{(27 - 8 z^2)}{6} \AH{2}
                       - \frac{( 1 - z^2)}{ z} \AH{1}^2  
                       - \frac{(1 - z^4) (3 - 26 z^2 + 3 z^4)}{32 z^4} \AH{1} \SH{1}
                       \nonumber\\
                       &
                       - \frac{(1 - z^4) (1 - 4 z^2 + z^4)}{ 4 z^4} \AH{1}\log{2}
                        + 2 \AH{2} \SH{1} 
                        + \frac{(1 - 14 z^2 + z^4)}{4 z^2} (\AH{2, -1} + 3 \AH{2, 1}) 
                        \nonumber\\
                       &
                       + \frac{(1 - 6 z^2 + z^4)}{2 z^2 } \left(\AH{2}\log{2}  + \AH{3}\right) \left.+ \frac{(1 - z^2)^5}{64 z^5} \AH{1} \left(4 \AH{1} \log{2}+ \AH{1} \SH{1} - 2 \AH{1, -1} \right)
                         \right)
\end{align}
where $\AH{\boldsymbol a} = H_{\boldsymbol a}(z) - H_{\boldsymbol a}(-z)$, $\SH{\boldsymbol a} = H_{\boldsymbol a}(z) + H_{\boldsymbol a}(-z)$, $z=\frac{\sqrt{\en-m}}{\sqrt{\en+m}}$ and  $H_{\boldsymbol a}(z)=H_{a_1,\dots,a_n}(z)$ denotes a harmonic polylogarithm (HPL)~\cite{Remiddi:1999ew}. All HPLs have the transcendental weight 3, at most, and can be expressed in terms of classical polylogarithms $\Li_2$ and $\Li_3$ with complicated argument, see Appendix~\ref{sec:phiLiRes}.

\subsection*{Asymptotics}
The non-relativistic and high-energy asymptotics of Eq.~\eqref{eq:asyNLOAS} can
be compared with the corresponding results known from the literature. In particular, the
non-relativistic limit of the Bremsstrahlung spectrum is known since
Sommerfeld's paper \cite{sommerfeld1931beugung}. This formula is exact in the
parameters $Z\alpha/v,\ Z\alpha/v^\prime$, where $v$ and $v^\prime$ are the
velocities of the initial and final electron. In contrast, our perturbative
result is valid for any $v, v^\prime$, but implies that $Z\alpha$ is the
smallest parameter, i.e., that $Z\alpha/v\ll 1$. Therefore, we should expand the
Sommerfeld spectrum in $Z\alpha/v$ and $Z\alpha/v^\prime$, multiply it by
$\omega/\en$, and integrate over $\omega$ from zero to $mv^2/2$. For the contribution $\propto (Z\alpha)^3$ we obtain
\begin{equation}
  \label{eq:phiAsyNR}
  \bar{\phi}_{\mathrm{NLO}} \stackrel{\en \to  m}{\sim}  \frac{16\pi}{3v}(2 \log{2}-1)\,
\end{equation}
which precisely reproduces the asymptotics of our result.

The high-energy limit of our result has the form
\begin{equation}
  \label{eq:phiAsyUR}
  \bar{\phi}_{\mathrm{NLO}} \stackrel{\en\gg m}{\sim} \frac{\pi}{24}\left( 19\pi^2 + 84(2-\zeta_3) - 60 \pi^2 \log{2} \right) +  \boxed{\frac{\pi^3}{2}} \log{\frac{\en}{m}}.
\end{equation}

In Ref.~\cite{Lee2005} the spectrum of the high-energy Bremsstrahlung has been calculated exactly in $Z\alpha$. The formula for the spectrum obtained in this Ref. implies that the final electron remains ultra-relativistic. Meanwhile, it is easy to establish that the region $\en^\prime\sim m$ contribute to the non-logarithmic constant in Eq. \eqref{eq:phiAsyUR}. Thus, we can only compare the coefficient in front of the logarithm, boxed in  Eq. \eqref{eq:phiAsyUR}, which perfectly agrees. Note that the end of the spectrum has been considered in Ref. \cite{di2010high}, mostly motivated by applications to heavy atoms, $Z\alpha\sim 1$. The result of this paper has the form of slowly convergent sum of nested integrals which is difficult to calculate with sufficient precision in our present restrictions $Z\alpha\ll 1$.

Although our present consideration does not allow to obtain photon spectrum, we
can use the calculated master integrals to obtain result for the cross section
integrated with the weight $\omega^k$ for $k=1,2,\ldots$.
\footnote{We get the first six moments ($n=0,\ldots,5$) by simply making the IBP reduction of the cross section weighted with $\omega^n=D_{12}^n$. For higher moments the complexity of the IBP reduction gradually increases and we apply instead an approach described in Section~\ref{sec:fitspec}.}
In particular, we can compare the high- and low-energy asymptotics these quantities with the corresponding results obtained by integration of the high- and low-energy spectra. It is convenient to consider the following quantity
\begin{equation}
    \label{eq:ewCS}
    K^{(n)} (Z,\en)= \sigma_0\left[K^{(n)}_{\mathrm{LO}}+(Z\alpha)K^{(n)}_{\mathrm{NLO}}+\ldots \right]\stackrel{\text{def}}{=}\int\limits_{0}^{T} \left(1 - {\omega}/{T} \right)^n \frac{\omega}{\en} d\sigma_{eZ\to e\gamma Z}(Z,\en),
\end{equation}
where $T=\en - m$ is the kinetic energy of the incoming electron.
From Ref. \cite{Lee2005}, by direct integration of the spectrum we obtain for $n>0$
\begin{equation}
    \label{eq:KnZal}
    K^{(n)}_{\mathrm{NLO}} = \frac{\pi^3 (3 + 2 n) (4 + 9 n + 3 n^2)}{4 n (1 + n) (2 + n) (3 + n)}\frac{m}{\en} + \mathcal{O}\left( \frac{1}{\en^2} \right)\,.
\end{equation}
Note that the contribution of the high-frequency end of the spectrum is negligible for positive $n$. 
Expanding our results for $K^{(n)}_{Z\alpha}$ in the limit $\en\to\infty$ we find exact agreement with Eq.~\eqref{eq:KnZal} for $n=1,\ldots, \nmax$. This comparison is a strong check on the validity of the procedure, calculated master integrals and performed IBP reduction.

Similarly, we can compare the low-energy asymptotics of $K^{(n)}$ with the corresponding results obtained by a straightforward integration of the Sommerfeld formula expansion,
\begin{equation}
    K^{(n)}_{\mathrm{NLO}} = \frac{8\pi  }{3v} \left[\frac{(4 n+3) S_1(n)}{(n+1) (2 n+1)}-\frac{2 S_1(2 n+1)}{n+1}+\frac{4 \log 2}{2 n+1}\right] +  \mathcal{O}\left(v\right)\,,
\end{equation} 
where $S_1(n)=\sum_{j=1}^{n}\frac1{j}$. Again, we find perfect agreement for $n=0,\ldots, \nmax$.

\subsection*{Numerical results and plots}

\begin{figure}[t]
    \centering
    \includegraphics[width=\textwidth]{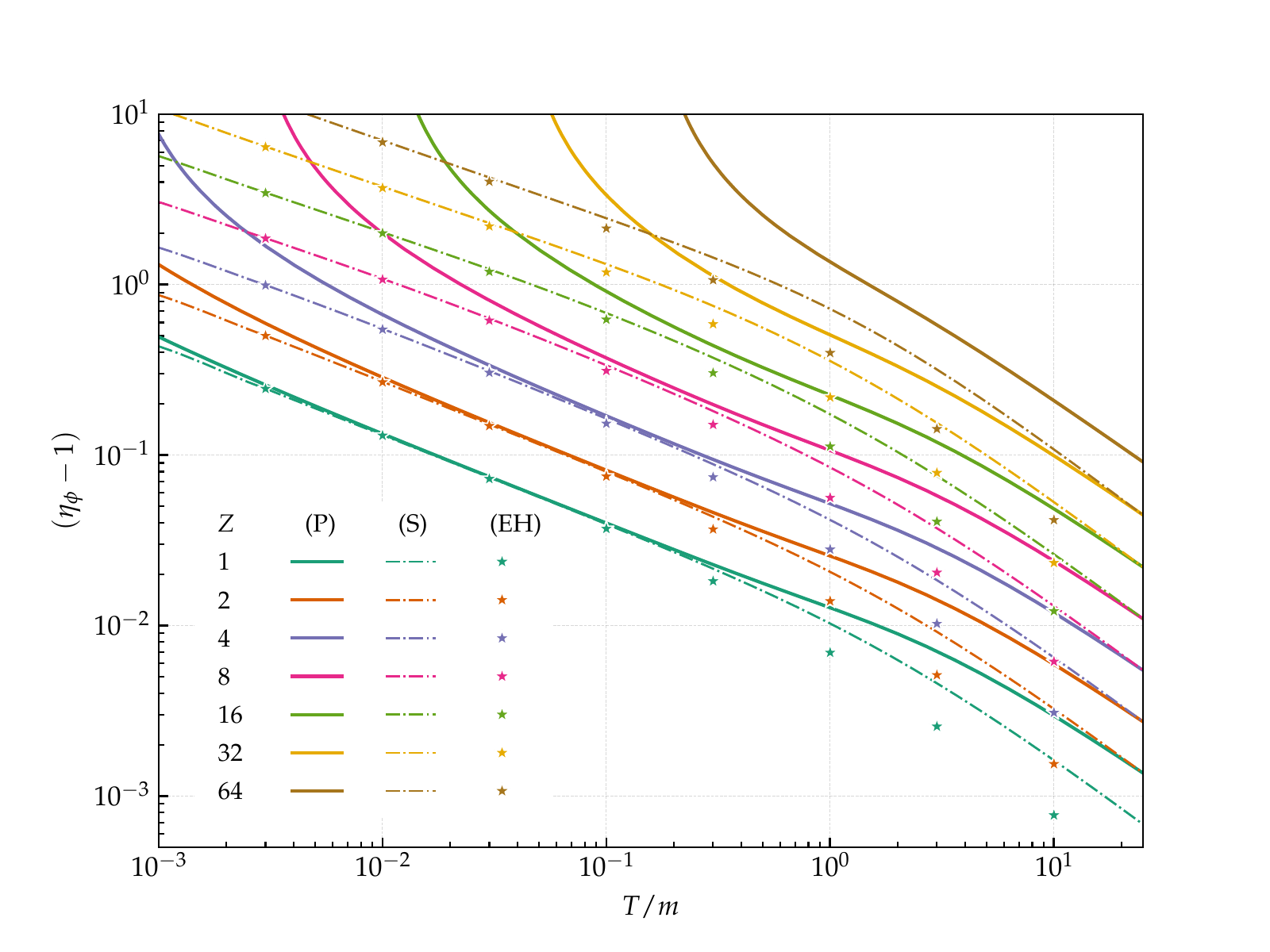}
    \caption{Comparison of the electron to positron energy loss ratio
        $\eta_{\phi}$ for different $Z$ values. Here (P) is a perturbative NLO
        result from the current paper and (S) is a non-relativistic exact to all orders
        in $Z\alpha/v$ result by Sommerfeld~\cite{sommerfeld1931beugung}, (EH) is a result of Elwert-Haug~\cite{PhysRev.183.90}.}
    \label{fig:asy-compare}
\end{figure}

For the sake of comparison of our results with those available in the literature, we present in Fig. \ref{fig:asy-compare} the electron-positron asymmetry in energy loss defined as
\begin{equation}
  \label{eq:etaPhiDef}
  \eta_{\phi} = \frac{\phi(+Z,\en)}{\phi(-Z,\en)}
\end{equation}
as the function of energy. Together with our perturbative results we plot those obtained using the Sommerfeld spectrum~\cite{sommerfeld1931beugung} and Elwert-Haug interpolation~\cite{PhysRev.183.90}.

In order to obtain Sommerfeld and Elwert-Haug cross sections it was crucial to have fast high-precision numerical evaluation of the Gauss hypergeometric function with complex parameters. In our work we use the implementation provided by the multiprecision floating point library \texttt{Arb} \cite{Johansson2017arb}. For the Elwert-Haug cross section we perform in addition the  four-dimensional integration of the EH cross section with \texttt{Cuba} library~\cite{Hahn:2016ktb}.
Of course, since our result is expressed via classical polylogarithms, its numerical evaluation is straightforward.

The deviation of Sommerfeld result from our result at small enough $T$ even for small $Z$ is not surprising. As we already mentioned our formula does not take into account the corrections of relative order $(Z\alpha/v)^n$ for $n\geqslant 2$, thus, when $2T\lesssim (Z\alpha)^2$, Sommerfeld result is reliable, while ours expectedly gradually looses its validity.

The deviation of Elwert-Haug result from ours even for small $Z$ at high enough energies can be explained as follows. The original goal of Ref.~\cite{PhysRev.183.90} was not to correctly describe the asymmetry, but rather to interpolate between two asymptotics known at that time: the Sommerfeld result and the Bether-Heitler high-energy asymptotics. The latter is symmetric with respect to the change $Z\to -Z$, so the Elwert-Haug asymmetry falls off faster than the Sommefeld's. It is interesting to note that our result for the asymmetry goes above the Sommefeld's.

\begin{figure}[t]
  \centering
  \includegraphics[width=\textwidth]{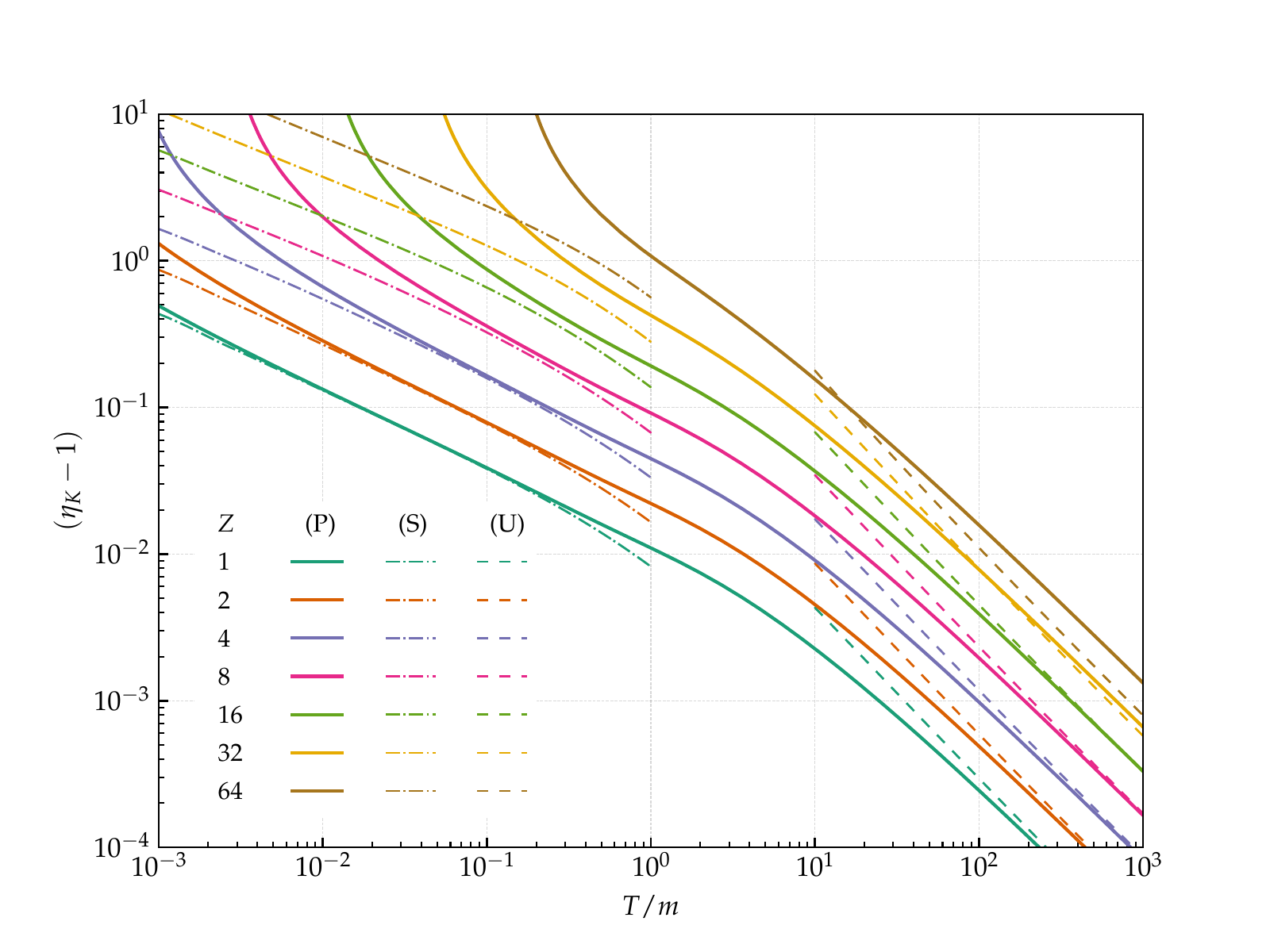}
  \caption{Energy weighted asymmetry, (P) - perturbative NLO result, (S) -
    Sommerfeld \cite{sommerfeld1931beugung}, (U) - corrections from \cite{Lee2005}}
  \label{fig:asyE-compare}
\end{figure}

In Ref. \cite{Lee2005} the first correction to the Bethe-Maximon spectrum has been calculated. This correction is antisymmetric in $Z$, and thus defines the charge asymmetry in the Bremsstrahlung spectrum. However, as explained above, when it concerns energy loss, the spectrum of Ref. \cite{Lee2005} can be used to obtain it only with logarithmic accuracy, and the extraction of the additive constant from the results of Ref. \cite{di2010high} appears to be not practically possible.  Due to these reasons, instead of $\eta_{\phi}$ we use a related asymmetry,
\begin{equation}
    \label{eq:etaKDef}
    \eta_{K} = \frac{K^{(1)}(+Z,\en)}{K^{(1)}(-Z,\en)}
\end{equation}
where $K^{(n)}$ is defined in Eq. \eqref{eq:ewCS}. This is asymmetry in the cross section integrated with the weight $\omega\en^\prime$ (to be compared with weight $\omega$ for $\phi$).
In Fig.~\ref{fig:asyE-compare} we provide a comparison of the asymmetry between positron and electron energy-weighted energy-loss functions $K^{(1)}$ with results for non-relativistic evaluation by Sommerfeld and ultra-relativistic limit predictions from~\cite{Lee2005}. Our result agrees with the latter at small $Z$ and expectedly fails when $Z$ becomes large.

\subsection*{Fitting spectrum from fixed moments}
\label{sec:fitspec}
As we already mentioned, our approach is suitable for the calculation of the moments, i.e.,
integrals of the cross section weighted with $\omega^n$, or, equivalently,
$K^{(n)}$ defined in Eq. \eqref{eq:ewCS}. 

Although the computational complexity
of the IBP reduction gradually grows with $n$, we have managed to obtain
$K^{(n)}_{\mathrm{NLO}}$ for $n=0,\ldots,5$ rather easily with direct
application of IBP reduction tables. In order to obatin yet more moments, we have use the approach based on the derivation of the system of difference equations in variable $n$. To construct this
system, we use the fact that  the integrals of the family \eqref{eq:jIntPows} which appear in the different moments are the same except the last index $n_{12}$ since $D_{12}=(k\cdot n)=\omega$. 
Since
\begin{equation}
  \label{eq:auxMassProp}
  \frac{1}{1-t (k\cdot n)} = \sum\limits_{i=0}^{\infty} (k\cdot n)^i t^i ,
\end{equation}
we can obtain the generating function for the moments by replacing $D_{12}$ with $1-t (k\cdot n)$.
Constructing the system of differential equations for integrals with $t$-dependent
propagators~\eqref{eq:auxMassProp} with appropriate ansatz in the form of the
series $J=\sum c_i t^i$ we obtain recurrence relations for the coefficients
$c_i$ which is nothing but the corresponding moments of original integrals
\eqref{eq:jIntPows}.
On this way, we have calculated moments up to $n=\nmax$. This has provided a number of checks for low- and high-energy asymptotics as explained previously.

Besides, having a considerable number of moments at hand, it is tempting to try to recover the NLO photon spectrum for arbitrary energies.
It is convenient to make a fit for the following quantity
\begin{equation}
    \spe(z,\tau)=(Z\alpha \sigma_0)^{-1}p \frac{\omega}{\en}\frac{d\sigma_{\mathrm{NLO}}}{d\tau}\,,
\end{equation}
 where $\tau=\sqrt{T^\prime/T}=\sqrt{\frac{\en^\prime - m}{\en - m}}$. The convenience of this quantity is that it has finite low- and high- energy limits, known thanks to Refs. \cite{sommerfeld1931beugung} and \cite{Lee2005}, respectively. They read
\begin{align}
  \spe(0,\tau) &=\frac{16\pi  }{3} (1-\tau) \log \left(\frac{1+\tau }{1-\tau}\right)\,,\label{eq:zLowSig}\\
  \spe(1,\tau) &=\frac{\pi ^3 }{2 \tau }\left(\tau ^2+1\right) \left(2 \tau ^4-\tau ^2+2\right)\,.\label{eq:zHighSig}
\end{align}
We tried several fitting bases and found that $\{1,\tau,\ldots, \tau^{7}\}$
gives the most stable results\footnote{Inclusion of yet higher moments into the fit seems to  introduce some artifacts in the spectrum.}. In other words, we write $\spe(z,\tau)=\sum_{m=0}^{7}C_m(z) \tau^m$ and fit coefficients to reproduce the integrals $K^{(n)}_{\mathrm{NLO}}$ with $n=0,\ldots, 7$. Given that $pK^{(n)}_{\mathrm{NLO}}(z)=\int_{0}^{1}d\tau \spe(z,\tau)\tau^{2n}$, we can write the coefficients as
 \begin{equation}
     C_m(z)=\frac{2z}{1-z^2} \sum_{n=0}^{7}\left(T^{-1}\right)_{mn}K^{(n)}_{\mathrm{NLO}},
 \end{equation}  
 where $T=\{T_{nm}\}=\{(2n+m+1)^{-1}\}$.
 
\begin{figure}
    \includegraphics[width=\textwidth]{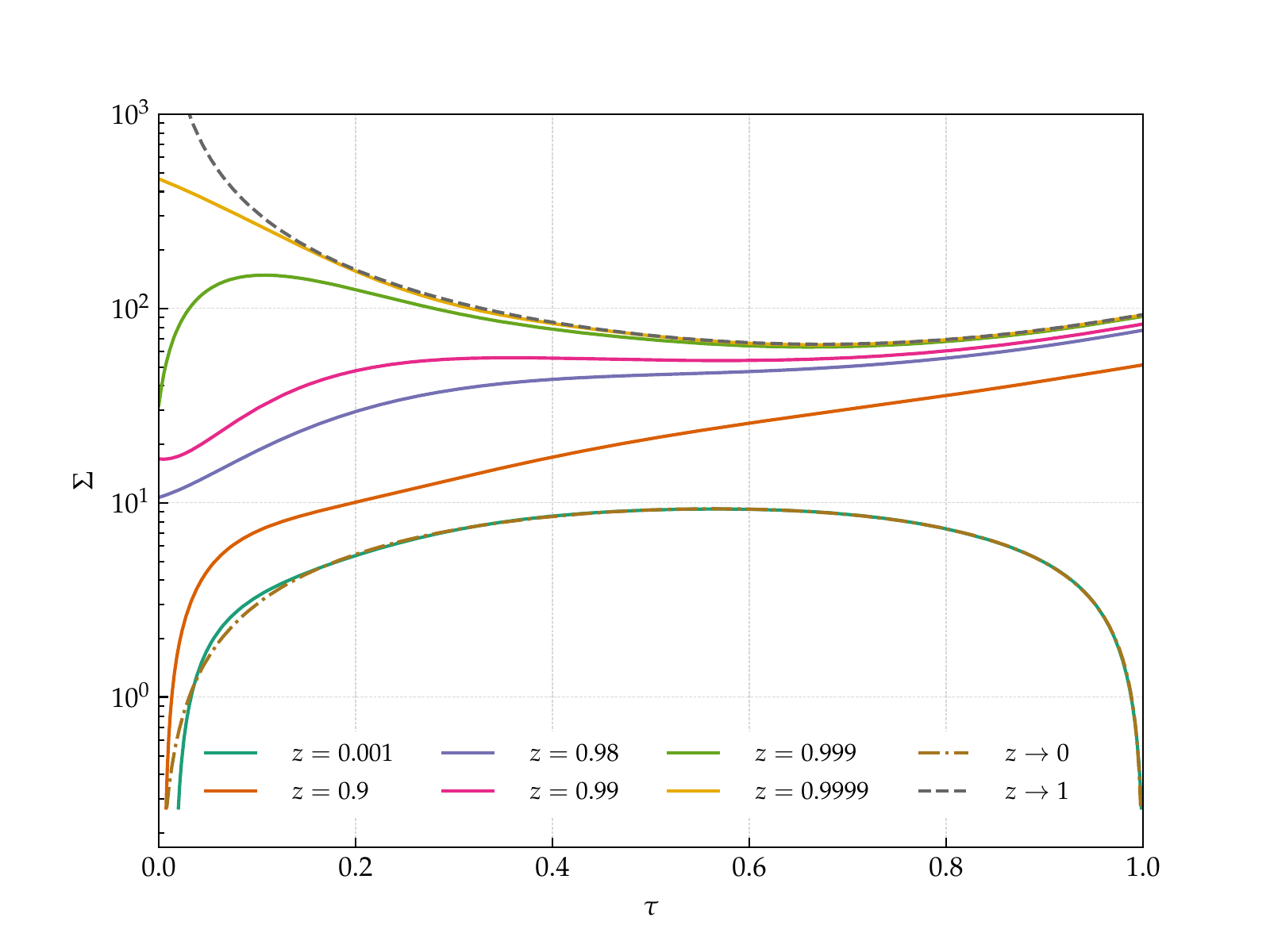}
    \caption{Function $\spe(z,\tau)$ for several values of $z=\frac{\sqrt{\en-m}}{\sqrt{\en+m}}$ and its comparison
    with low energy (dash-dotted curve) and high energy (dashed curve) asymptotics.}\label{fig:spectrum}
\end{figure}

The resulting spectra are shown in Fig. \ref{fig:spectrum}. 
 
\section{Conclusion}
\label{sec:conclusion}

To summarize, we have calculated the first Coulomb correction to the energy loss of electron due to Bremsstrahlung in the field of a nucleus for arbitrary electron energy. This correction leads to the asymmetry in the energy loss of an electron and a positron. We have compared the high- and the low-energy asymptotics with available asymptotics and found a perfect agreement. We have calculated also a number of moments of the cross section, as defined in Eq. \eqref{eq:ewCS}. The low- and high-energy asymptotics of these moments perfectly agrees with the known results which provides a stringent test of the consistency of our approach. Besides, we have used the calculated moments to approximately recover the photon spectrum, which might be interesting from the point of view of some applications. It would be interesting to compare our fit with direct calculation of the spectrum once/if the latter will be available at some point.

From the perspective of developing new methods, we have demonstrated how to obtain the dimensional recurrence relations for the boundary constants starting from the equations for original master integrals and used this technique to apply the DRA method \cite{DRA2010} to the calculation of the constants.


\acknowledgments
This work was supported by Russian Science Foundation, grant 20-12-00205. R.L. is grateful to A.I. Milstein for useful discussions.

\appendix

\section{$\AH{}$ and $\SH{}$ functions via classical polylogarithms}
\label{sec:phiLiRes}

\begin{align}
  \label{eq:AStoLi}
  \SH{1} & = -\log(1 - z) - \log(1 + z)\\
  \AH{1} & = -\log(1 - z) + \log(1 + z)\,,\\
  \AH{2} & = \Li_2(z) - \Li_2(-z)\,,\\
  \AH{3} & = \Li_3(z) - \Li_3(-z)\,,\\
  \AH{1,-1} & = \left(  \log(1 + z) - \log(1 - z) \right) \log{2} 
              + \Li_2 \left( \frac{1 - z}{2} \right)
              - \Li_2\left( \frac{1 + z}{2} \right)\,,\\
  \AH{2,1} & = \frac{1}{6}\left(  \log^3(1 + z) - \log^3(1 - z) \right)
             - \Li_2(z) \log(1 - z) + \Li_2(-z) \log(1 + z)   \nonumber\\
         & + \Li_3(z) - \Li_3(-z) 
           - \Li_3\left( \frac{z}{1 + z}\right) +\Li_3\left( \frac{-z}{1 - z} \right) \,,\\
  \AH{2,-1} & =\left( \frac{\pi^2}{12} - \frac{\log^2{2}}{2}  \right) \left( \log(1 + z)-\log(1 - z) \right)\nonumber\\
         & + \left( \log^2(1 + z) -  \log^2(1 - z) \right) \frac{\log{2}}{2}
         +  \Li_2( z) \log(1 + z) -  \Li_2( -z)\log(1 - z)
           \nonumber\\
         & + \Li_3\left( \frac{1 - z}{2} \right) - \Li_3\left( \frac{1+z}{2}\right)- \Li_3\left(\frac{-2 z}{1 - z} \right) + \Li_3\left(\frac{2 z}{1 + z}\right)
            \nonumber\\
         &  - \Li_3( z)+ \Li_3( -z)
            - \Li_3\left( \frac{z}{1 + z}\right) + \Li_3\left( \frac{-z}{1 - z} \right)\,.
\end{align}

\bibliographystyle{JHEP}
\bibliography{BSA}

\providecommand{\href}[2]{#2}\begingroup\raggedright\begin{thebibliography}{10}

\bibitem{heitler1933stopping}
W.~Heitler and F.~Sauter, \emph{Stopping of fast particles with emission of
  radiation and the birth of positive electrons},
  \href{https://doi.org/10.1038/132892a0}{\emph{Nature} {\bfseries 132} (1933)
  892}.

\bibitem{BetheHeitler1934}
H.~Bethe and W.~Heitler, \emph{On the stopping of fast particles and on the
  creation of positive electrons},
  \href{https://doi.org/10.1098/rspa.1934.0140}{\emph{Proc.Roy.Soc.Lond.}
  {\bfseries 146} (1934) 83}.

\bibitem{Racah1934}
G.~Racah, \emph{Sopra l’irradiazione nell’urto di particelle veloci},
  \href{https://doi.org/10.1007/BF02959918}{\emph{Il Nuovo Cimento} {\bfseries
  11} (1934) 461}.

\bibitem{BethMax1954}
H.~A. Bethe and L.~C. Maximon, \emph{Theory of bremsstrahlung and pair
  production. i. differential cross section},
  \href{https://doi.org/10.1103/PhysRev.93.768}{\emph{Phys. Rev.} {\bfseries
  93} (1954) 768}.

\bibitem{OlseMax1959}
H.~Olsen and L.~C. Maximon, \emph{Photon and {E}lectron {P}olarization in
  {H}igh-{E}nergy {B}remsstrahlung and {P}air {P}roduction with {S}creening},
  \href{https://doi.org/10.1103/PhysRev.114.887}{\emph{Phys.~{R}ev.} {\bfseries
  114} (1959) 887}.

\bibitem{Lee2005}
R.~N. Lee, A.~I. Milstein, V.~M. Strakhovenko and O.~Y. Schwarz, \emph{Coulomb
  corrections to bremsstrahlung in electric field of heavy atom at high
  energies}, \href{https://doi.org/10.1134/1.1866193}{\emph{J. Exp. Theor.
  Phys.} {\bfseries 100} (2005) 1}
  [\href{https://arxiv.org/abs/hep-ph/0404224}{{\ttfamily hep-ph/0404224}}].

\bibitem{LiteRed2012}
R.~N. Lee, \emph{Presenting litered: a tool for the loop integrals reduction},
  2012.

\bibitem{LiteRed2013}
R.~N. Lee, \emph{{LiteRed 1.4: a powerful tool for reduction of multiloop
  integrals}}, \href{https://doi.org/10.1088/1742-6596/523/1/012059}{\emph{J.
  Phys. Conf. Ser.} {\bfseries 523} (2014) 012059}
  [\href{https://arxiv.org/abs/1310.1145}{{\ttfamily 1310.1145}}].

\bibitem{LiteRed2021}
R.~N. Lee, ``Litered2.'' to be published soon.

\bibitem{Henn2013}
J.~M. Henn, \emph{{Multiloop integrals in dimensional regularization made
  simple}},
  \href{https://doi.org/10.1103/PhysRevLett.110.251601}{\emph{Phys.Rev.Lett.}
  {\bfseries 110} (2013) 251601}
  [\href{https://arxiv.org/abs/1304.1806}{{\ttfamily 1304.1806}}].

\bibitem{Lee2014}
R.~N. Lee, \emph{{Reducing differential equations for multiloop master
  integrals}}, \href{https://doi.org/10.1007/JHEP04(2015)108}{\emph{J. High
  Energy Phys.} {\bfseries 1504} (2015) 108}
  [\href{https://arxiv.org/abs/1411.0911}{{\ttfamily 1411.0911}}].

\bibitem{Libra2020}
R.~N. Lee, \emph{Libra: A package for transformation of differential systems
  for multiloop integrals},
  \href{https://doi.org/https://doi.org/10.1016/j.cpc.2021.108058}{\emph{Computer
  Physics Communications} {\bfseries 267} (2021) 108058}.

\bibitem{Lee:2017qql}
R.~N. Lee, A.~V. Smirnov and V.~A. Smirnov, \emph{{Solving differential
  equations for Feynman integrals by expansions near singular points}},
  \href{https://doi.org/10.1007/JHEP03(2018)008}{\emph{JHEP} {\bfseries 03}
  (2018) 008} [\href{https://arxiv.org/abs/1709.07525}{{\ttfamily
  1709.07525}}].

\bibitem{Lee:2019zop}
R.~N. Lee, A.~V. Smirnov, V.~A. Smirnov and M.~Steinhauser, \emph{{Four-loop
  quark form factor with quartic fundamental colour factor}},
  \href{https://doi.org/10.1007/JHEP02(2019)172}{\emph{JHEP} {\bfseries 02}
  (2019) 172} [\href{https://arxiv.org/abs/1901.02898}{{\ttfamily
  1901.02898}}].

\bibitem{DRA2010}
R.~Lee, \emph{Space-time dimensionality d as complex variable: Calculating loop
  integrals using dimensional recurrence relation and analytical properties
  with respect to d}, \href{https://doi.org/DOI:
  10.1016/j.nuclphysb.2009.12.025}{\emph{Nucl. Phys. B} {\bfseries 830} (2010)
  474} [\href{https://arxiv.org/abs/0911.0252}{{\ttfamily 0911.0252}}].

\bibitem{SummerTime2016}
R.~N. Lee and K.~T. Mingulov, \emph{{Introducing SummerTime: a package for
  high-precision computation of sums appearing in DRA method}},
  \href{https://doi.org/10.1016/j.cpc.2016.02.018}{\emph{Comput. Phys. Commun.}
  {\bfseries 203} (2016) 255}
  [\href{https://arxiv.org/abs/1507.04256}{{\ttfamily 1507.04256}}].

\bibitem{PSLQ}
H.~Ferguson, D.~Bailey and S.~Arno, \emph{Analysis of pslq, an integer relation
  finding algorithm}, {\emph{Mathematics of Computation} {\bfseries 68} (1999)
  351}.

\bibitem{Remiddi:1999ew}
E.~Remiddi and J.~A.~M. Vermaseren, \emph{{Harmonic polylogarithms}},
  \href{https://doi.org/10.1142/S0217751X00000367}{\emph{Int. J. Mod. Phys. A}
  {\bfseries 15} (2000) 725}
  [\href{https://arxiv.org/abs/hep-ph/9905237}{{\ttfamily hep-ph/9905237}}].

\bibitem{sommerfeld1931beugung}
A.~Sommerfeld, \emph{Über die beugung und bremsung der elektronen},
  \href{https://doi.org/10.1002/andp.19314030302}{\emph{Annalen der Physik}
  {\bfseries 403} (1931) 257}.

\bibitem{di2010high}
A.~Di~Piazza and A.~Milstein, \emph{High-energy electron-positron
  photoproduction cross section close to the end of the spectrum},
  {\emph{Physical Review A} {\bfseries 82} (2010) 042106}.

\bibitem{PhysRev.183.90}
G.~Elwert and E.~Haug, \emph{Calculation of bremsstrahlung cross sections with
  sommerfeld-maue eigenfunctions},
  \href{https://doi.org/10.1103/PhysRev.183.90}{\emph{Phys. Rev.} {\bfseries
  183} (1969) 90}.

\bibitem{Johansson2017arb}
F.~Johansson, \emph{Arb: efficient arbitrary-precision midpoint-radius interval
  arithmetic}, \href{https://doi.org/10.1109/TC.2017.2690633}{\emph{IEEE
  Transactions on Computers} {\bfseries 66} (2017) 1281}.

\bibitem{Hahn:2016ktb}
T.~Hahn, \emph{{Concurrent Cuba}},
  \href{https://doi.org/10.1016/j.cpc.2016.05.012}{\emph{Comput. Phys. Commun.}
  {\bfseries 207} (2016) 341}.

\end{thebibliography}\endgroup
\end{document}